\documentclass[preprint,aps,superscriptaddress]{revtex4}

\usepackage{graphicx}% Include figure files
\usepackage{dcolumn}% Align table columns on decimal point
\usepackage{bm}% bold math
\usepackage{natbib}% bold math
\begin{document}

\preprint{APS/123-QED}

\title{In-situ NMR Measurements of Vapor Deposited Ice.}

%\author{M. Diamant}
% \affiliation[Also at ]{Schulich Faculty of Chemistry
%Technion - Israel Institute of Technology
%Technion City, Haifa 32000, Israel.}%Lines break automatically or can be forced with \\
%\author{S. Rahav}%
%\affiliation{Schulich Faculty of Chemistry
%Technion - Israel Institute of Technology
%Technion City, Haifa 32000, Israel.}
%\author{R. Ferrando}
% \affiliation[ Dipartimento di Fisica, Universita' di Genova - v. Dodecaneso 33 - 16146, Genova, Italy }%Lines break %automatically or can be forced with \\
%\author{G. Alexandrowicz}%
%\affiliation{Schulich Faculty of Chemistry
%Technion - Israel Institute of Technology
%Technion City, Haifa 32000, Israel.}
%\email{ga232@tx.technion.ac.il}

\author{\textbf{E. Lisitsin-Baranovsky}}

\affiliation{\textbf{Schulich Faculty of Chemistry, Technion - Israel Institute of Technology, Technion City, Haifa 32000, Israel.}}

\author{\textbf{S. Delage}}

\affiliation{\textbf{{D\'epartement de Chimie, Universit\'e  de Sherbrooke, 2500 Boulevard Universit\'e, Sherbrooke, Qu\'ebec, Canada J1K 2R1}}}

\author{\textbf{O. Sucre}}

\affiliation{\textbf{Schulich Faculty of Chemistry, Technion - Israel Institute of Technology, Technion City, Haifa 32000, Israel.}}

\author{\textbf{O. Ofer}}

\affiliation{\textbf{Schulich Faculty of Chemistry, Technion - Israel Institute of Technology, Technion City, Haifa 32000, Israel.}}

\author{\textbf{P. Ayotte}}

\affiliation{\textbf{{D\'epartement de Chimie, Universit\'e  de Sherbrooke, 2500 Boulevard Universit\'e, Sherbrooke, Qu\'ebec, Canada J1K 2R1}}}

\author{\textbf{G. Alexandrowicz}}

\email{ga232@tx.technion.ac.il  Tel: +97248295563}

\affiliation{\textbf{Schulich Faculty of Chemistry, Technion - Israel Institute of Technology, Technion City, Haifa 32000, Israel.}}
%\selectlanguage{english}

\begin{abstract}
In-situ NMR spin-lattice relaxation measurements were performed on several vapor deposited ices. The measurements, which span more than 6 orders of magnitude in relaxation times, show a complex spin-lattice relaxation pattern that is strongly dependent on the growth conditions of the sample. The relaxation patterns change from multi-timescale relaxation for samples grown at temperatures below the amorphous-crystalline transition temperature to single exponential recovery for samples grown above the transition temperature. The slow-relaxation contribution seen in cold-grown samples exhibits a temperature dependence, and becomes even slower after the sample is annealed at 200K. The fast-relaxation contribution seen in these samples, does not seem to change or disappear even when heating to temperatures where the sample is evaporated. The possibility that the fast relaxation component is linked to the microporous structures in amorphous ice samples is further examined using an environmental electron scanning microscope. The images reveal complex meso-scale microporous structures which maintain their morphology up to their desorption temperatures. These findings, support the possibility that water molecules at pore surfaces might be responsible for the fast-relaxation contribution. Furthermore, the results of this study indicate that the pore-collapse dynamics observed in the past in amorphous ices using other experimental techniques, might be effectively inhibited in samples which are grown by relatively fast vapor deposition.
\end{abstract}

\pacs{Valid PACS appear here}% PACS, the Physics and Astronomy
                             % Classification Scheme.
%\keywords{Suggested keywords}%Use showkeys class option if keyword
                              %display desired
\maketitle

\section{Introduction}

Solid materials, grown from vapor deposition on a cold surface, are encountered in a wide range of research fields and applications. One example is the field of astrochemistry where amorphous solid water (ASW), as well as other molecular ices, play an important role in the formation of molecules in the interstellar medium \cite{Herbst1995}. A different example, is the field of atmospheric chemistry, where the solid particles grown from the vapor phase, participate in the reactions which control the composition of the atmosphere and ultimately Earth's climate \cite{MOLINA1253, TOLBERT1258}. In both of the above examples, the structure and morphology of these solids is determined by the growth conditions and has major implications on their functional properties and surface chemistry which can take place.
Nuclear Magnetic Resonance (NMR) is a powerful technique for studying the atomic scale structure and dynamics of materials. When studying delicate materials such as cryogenic molecular-solids, which are easily perturbed and modified by the irradiation with X-rays, visible and even infra-red radiation sources, the gentle nature of low-energy radio-frequency waves make NMR a particularly suitable choice. On the other hand, most NMR studies are performed using commercial NMR spectrometers, which can not be easily integrated with the apparatus used for vapor deposition of samples. While, techniques have been developed to transfer cryogenic solids from one setup to another, the question whether the samples change during these transfers leads to some uncertainty and makes it highly advantageous to perform in-situ studies of delicate vapor deposited cryogenic solids \cite{Loerting16}. 

One particularly important type of vapor deposited material is amorphous solid water (ASW), which is the term used to describe ice which condenses from the vapor phase onto a cold surface. ASW is the most abundant molecular solid in space and is believed to play a crucial role in the formation of planets\cite{Erenfreund,Watanabe2008}. While ASW was the first amorphous ice to be studied, understanding its properties and the nature of the transitions to and from other forms of ice are still an active research field with unanswered questions\cite{Loerting16}. 

The current understanding of the various forms of ice has relied on the combination of various complementary experimental techniques, and indeed NMR has contributed significantly to our understanding of crystalline ice\cite{petrenko2002physics}. Nevertheless, only a relatively small set of pioneering NMR experiments have been made on amorphous ices\cite{Ripmeester1992,Fujara2006,Fujara_PCCP_2013, Fujara2013}, and to the best of our knowledge NMR has not been used to study vapor deposited ices and ASW in particular, probably since this requires a rather non-conventional apparatus. In this manuscript, we describe an apparatus which allows growth of vapor deposited solids and NMR measurements without extracting the sample and modifying its properties. As a first application of the instrument, we studied the spin-lattice relaxation of vapor deposited ice, revealing an interesting multi-timescale relaxation process which is strongly linked to the deposition conditions.

\section{Apparatus and experimental details}

\begin{figure}
\includegraphics[width=170mm]{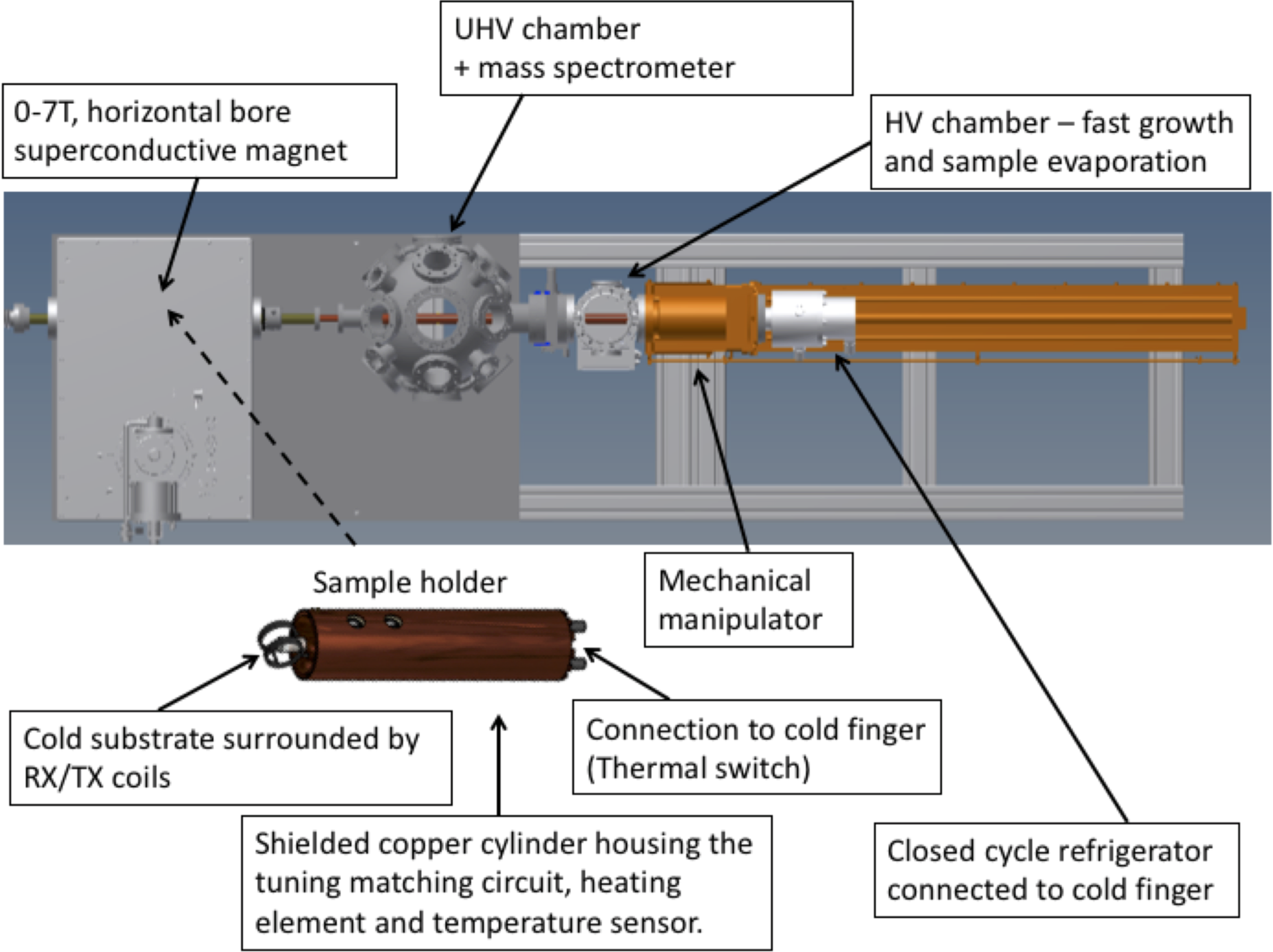}
%\begin{figure}
%\begin{center}\includegraphics[%
% width=85mm,
% keepaspectratio]{/Users/gilalexandrowicz/Desktop/simulation4_old/data/figure1_na115_small}\end{center}
%keepaspectratio]{figure1}\end{center}

%\begin{figure*}
%\begin{centering}
%\includegraphics[width=10cm]{figure1_na115_small}
%\par\end{centering}

\caption{\label{cap:Setup}Schematic drawing of the NMR apparatus (above) and the sample holder (below). }
\label{Setup}
\end{figure}

In order to grow and characterize solids using vapor deposition; Vacuum, gas-analysis, cryogenic and NMR technologies need to be combined. Figure \ref{Setup} shows a schematic of the apparatus we developed which combines all of these technologies, this setup is part of a larger apparatus we are currently developing, for measuring NMR from single surface layers, grown with a unique ortho-water molecular beam source \cite{Kravchuk319,Pierre12}. The main components which are indicated in the schematic are the horizontal-bore super-conducting variable-field magnet (0-7 T), into which the ultra high vacuum (UHV) titanium tube is inserted. The titanium tube is connected to a stainless steel UHV chamber which contains a quadrupole mass spectrometer (Hiden - HAL301). The UHV chamber is connected through a gate valve to a high vacuum (HV) chamber where relatively fast vapor deposition (up to $\approx 1$ mg/min) as well as sample evaporation can be performed without contaminating the UHV chamber. The sample itself is grown on a sample holder which is thermally connected, using a long copper rod (cold finger), to a commercial closed cycle refrigerator (Sumitomo – CH204). The sample can be moved between the center of the magnet, the spherical UHV chamber (where the mass spectrometer is located) and the HV part of the system using a motorized mechanical translator (McAllistair ZA4542). 

Figure 1 also shows a schematic of a sample holder. The sample holder consists of a cold substrate on which the sample is grown (a rod of Sapphire was used in the measurements presented below). The edge of the substrate is surrounded by copper coils, used for receiving and transmitting radio frequencies. The other side of the substrate is mounted to a copper cylinder which houses the capacitors for tuning and impedance matching, a heater and a temperature sensor (LakeShore, Cernox 1070) . The currently available temperature range is $30-380$K, where lower temperatures, down to 7K, should be possible using the same cooling system after installing additional radiation shielding. 
The apparatus includes two different sources for vapor deposition. One source is a highly collimated supersonic atomic/molecular beam which can be used to grow ultra-thin films. A second source, suitable for relatively large (mm$^3$) samples, was used in this study. It is located in the high vacuum chamber ($10^{-8}$ to $10^{-2}$ torr), and consists of a $100\mu$m laser-drilled aperture positioned on the end of a retractable gas line, connected on its other side to the vapor source. 

For the NMR measurements described below the vapor source was a glass ampule containing deionized water (18 M$\Omega -$ cm). The water was degassed by a combination of pumping and freeze thaw cycles. The mass spectrometer was used to monitor the deposition process and ensure that the degassing cycles removed the dissolved gasses effectively. Using the mass spectrometer, an upper limit of 0.1\% was determined for trace amounts of $O_2$, $N_2$, $CO$ and $CO_2$ . During the deposition, the water temperature was maintained at 0$^0$C to produce a stable vapor pressure behind the nozzle, whereas the nozzle was positioned approximately 5 mm from the substrate. A deposition rate of approximately 0.05 mg/min was calculated from the pressure increase in the high vacuum chamber and the known pumping speed. This rate is approximately equivalent \footnote{This value is rather crude and represents the average flux over the entire area of the substrate, since the samples grew into a cone shaped sample, the flux must be even higher at the center of the cone.} to a deposition flux of $2 \cdot 10^{17}$ cm$^{-2}$ sec$^{-1}$. After deposition, the source is retracted out of the way, allowing the sample to move freely into the UHV section. Finally, the in-vacuum NMR probe and tuning matching circuit were connected via a duplexer circuit and a low noise pre-amplifier (Miteq-AU-1114T) to a commercial NMR spectrometer (Redstone, Tecmag) and a power amplifier (TOMCO BT01000).

Complementary examinations of the morphology of vapor-deposited ice were performed using an environmental scanning electron microscope, (ESEM, Hitachi S-3000N, base pressure P$<10^{-5}$ torr). Ice films were synthesized in situ by vapor deposition on a sapphire plate mounted on a liquid nitrogen cooled cryo-stage (Gatan) housed in the ESEM chamber. The cryo-stage temperature was measured using a Pt resistance thermometer and controlled (in the 80K-200K) range by resistive heating using a PID controller. Ice films were grown by introducing a partial vapor pressure of H$_2$O (up to P $= 5\cdot10^{-3}$ Torr, for typically 20 minutes) in the ESEM chamber using a leak valve. Imaging of the samples was performed using secondary electrons detection and specimen charging was minimized using electron energies of at most 5kV. 

\section{Experimental results and discussion.}

Ice samples, studied by NMR, were grown at temperatures of 52K, 100K and 200K, two of which are well below the typical crystallization temperature (approx. $150K$\cite{Angell_review}) and one significantly above. Spin-lattice relaxation was measured at a magnetic field of 6.2 T using a saturation train followed by a variable time delay, $\tau_s$, and then by a solid echo ($90_x - \tau - 90_y$) sequence which produced the measured signal.

\begin{figure}
\includegraphics[width=170mm]{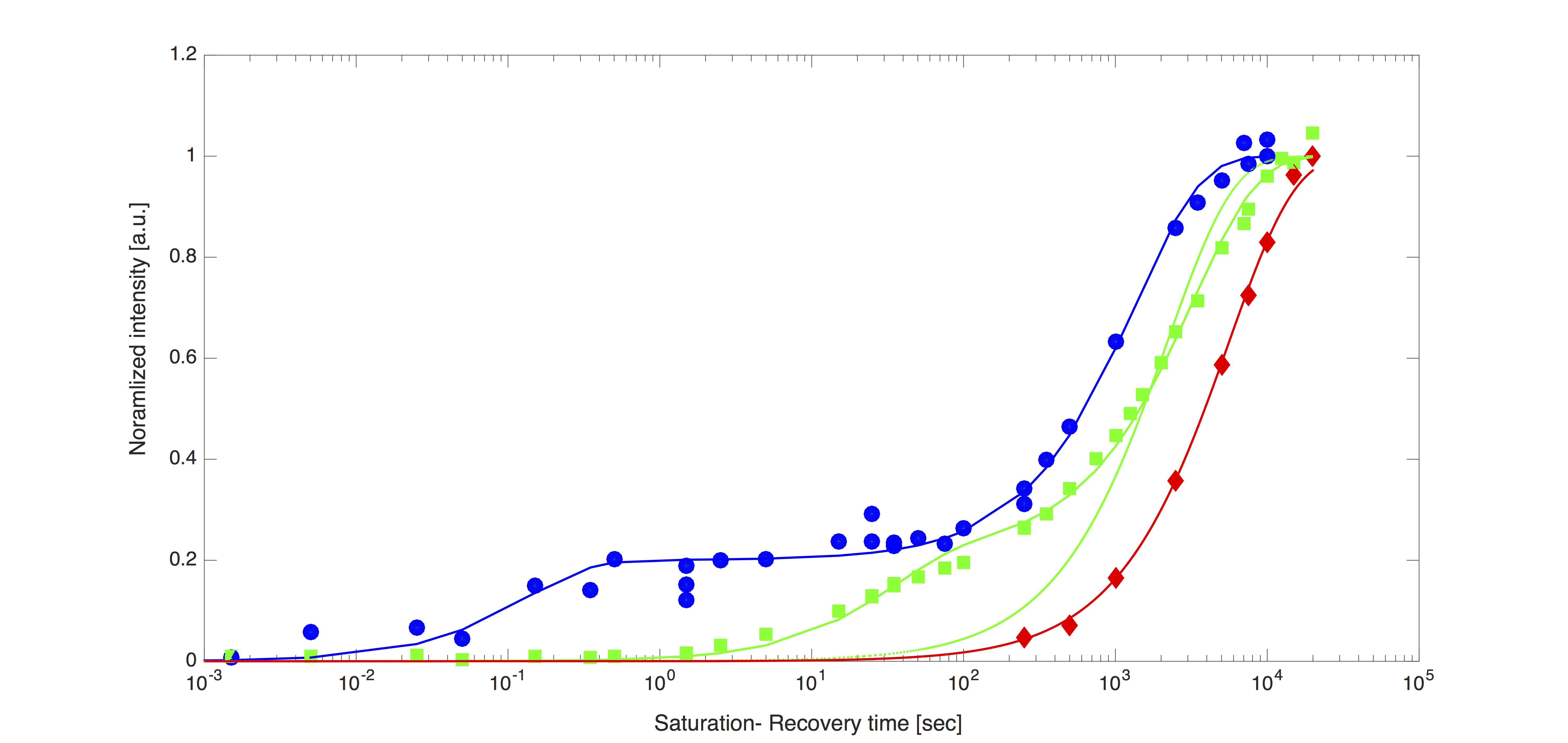}
\caption{\label{cap:Relax1}Spin-lattice relaxation curves for samples grown at different temperatures and measured at 100K. The circle (blue), square (green) and diamond (red) markers are for samples grown at 52K,100K and 200K respectively. The full lines are double (for the case of 52K and 100K measurements) and single (for the 200K data) exponential recovery fits. The dashed (green) line illustrates the failure of fitting the 100K data with a single exponential recovery model. }
\label{Relax1}
\end{figure}

Figure 2 shows the evolution of the signal as function of the delay time $\tau_s$, measured over more than 6 orders of magnitude. The blue circle markers are the results obtained for a sample which was grown at 52K and measured at 100K. A noticeable feature of the data is the appearance of two modes of relaxation, characterized by extremely
different time constants. A fraction of the signal recovers on a sub-second time scale, whereas the full signal recovers on a time scale which is 3 orders of magnitude slower. The blue line shows a double exponential fit to the data, $A_s (1- e^{-t/T_s})+A_L(1-e^{-t/T_L})$ yielding values of $T_s=0.1\pm{0.04}$ and $T_L=1350\pm{100}$ seconds for the short and long relaxation processes correspondingly. The square green markers show the results obtained for a sample which was grown at a hotter temperature of 100K, which is still significantly below the nominal crystallization temperature, and measured at 100K. While two distinct time scales for relaxation can still be seen, the fast relaxation is significantly slower than that of the 52K grown sample. The green line shows a fit to the same double exponential model used earlier ( $T_s=33\pm{4}$ sec and $T_L=3200\pm{150}$ sec ) which deviates systematically from the data, but still provides an approximate description of the dynamics. The dotted line shows an attempt to fit a single exponent to the longer relaxation process, shown to illustrate the inadequacy of single exponential models to fit the experimental curves, even when growing at 100K. Finally, the diamond red markers show the results obtained when the sample was grown at a high temperature (200K) and measured at 100K, here there seems to be only one, very slow relaxation process. The red line shows a single exponential model fit to the data yielding $T1=5600\pm{260}$sec.

The growth temperature affects the spin-lattice relaxation in two distinct ways, one is the appearance of a fast relaxing component at low temperatures, and the other is that the slow relaxing component becomes even slower as the growth temperature is increased. Previous measurements on pressure induced amorphous ice (LDA) have yielded relaxation rates which were one order of magnitude faster than those measured in a crystalline sample. Suggestions were made that this might reflect small amplitude motions such as librations or vibrations which are related to defects which exist in the amorphous state \cite{Ripmeester1992}. One obvious candidate for such defects in ASW are molecules in the vicinity of pores. ASW can be a highly porous solid, with a pore size distribution which strongly depends on the deposition conditions\cite{Mitterdorfer2014,Stevenson1999}. Pores in ASW are characterized by dangling OH bonds of molecules located close to the pores, hence, the different slow relaxation rates or the appearance of a fast relaxing component (or both phenomena) seen in figure 2, could perhaps be related to a high density of pores when growing at 52K, which reduces for higher growth temperatures of ASW and disappears when growing well above the crystallization temperature. This type of interpretation would also be consistent with fast relaxations observed in dielectric measurements of ASW which have been attributed to the enhanced rotational motion of water molecules located at the surface of the pores\cite{Johari1991}.

\begin{figure}
\includegraphics[width=170mm]{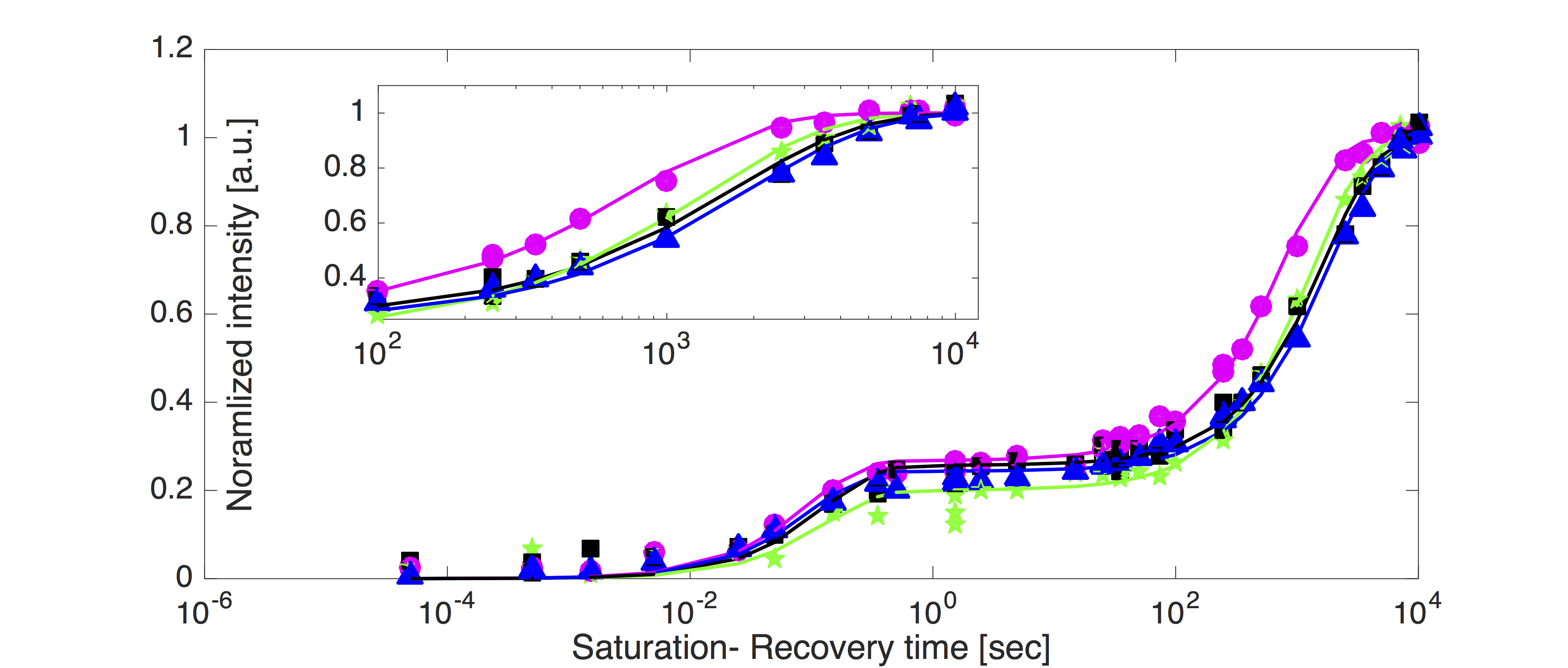}
\caption{\label{cap:Relax2}Spin-lattice relaxation curves of a sample grown at 52K and measured at 52K, 100K and 140K ( blue triangles, green stars and magenta circles respectively). The full lines with the matching colors are fits to the double exponential model presented in the text. Increased temperatures shorten the slow-relaxing component (seen more clearly in the inset which zooms on long recovery times), but do not produce a noticeable trend in the fast-relaxing component. Black squares show the results after annealing at 200K and cooling to 140K for the measurement. The annealing process elongates the slow-relaxing process, but does not seem to affect the fast relaxing component.}
\end{figure}

An additional set of experiments we performed is studying the evolution of the relaxation curves when heating the sample. The data presented in figure 3 shows the results for an ice sample grown at 52K, and measured for a series of temperatures 52K, 100K and 140K marked using triangles (blue) , stars (green), and circles (magenta), respectively. It can be seen that the slow relaxation component becomes gradually faster with increased temperature, reducing from $1940\pm{110}$ to $810\pm{40}$sec. Generally speaking, when increasing the temperature within this range, two opposite effects could be expected: one effect is sintering and annealing of defects , which according to the arguments mentioned above would be expected to elongate the relaxation times. A second effect is enhanced motion (diffusion, vibrations and librations), which in turn would be expected to shorten the relaxation times. From the measured trend of the slow relaxing component, it seems the dominant effect is the later. Interestingly, there is no significant change in the fast relaxation component within the accuracy of the experimental data, suggesting that either the two mechanisms mentioned above fortuitously balance each other, or that the origin of the fast relaxation mechanism is related to defects or motions which do not change much within this temperature range. 

Even more surprising are the results obtained after annealing the sample to 200K for 10 minutes and cooling back to 140K (black square markers in figure 3). On the one hand, the annealing has significantly elongated the slow relaxation component ($1700\pm{150}$ sec instead of $810$sec), making it more similar to that seen from hot-grown samples. On the other hand, the fast relaxing component does not seem to be affected. Thus, a film grown at a low temperature (52K or 100K) and annealed at 200K, does not resemble, in terms of its NMR spin-lattice relaxation, a sample which was grown at a higher temperature. Attempts to anneal at even higher temperatures lead to complete evaporation of the sample. \footnote{In addition to the temperature dependent effects described in the text, we noticed that individual samples grown in what should have been similar growth conditions, were characterized by different slow relaxation rates (varying by up to 70$\%$ in extreme cases), we believe this reflects a very strong dependency of the sample morphology on the exact (i.e., local) deposition conditions. Nevertheless all the temperature dependent trends discussed in the text (growth and measurement temperature) repeated themselves for all the samples we studied.}

If indeed the fast relaxation component in the data is due to defects in the structure of the films we grew, then these defects seem to be resilient to the rather extreme heating procedure we applied. The 200K annealing temperature is well above the nominal crystallization temperature. Furthermore, a few studies have shown that the porous structure collapses already at lower temperatures ($<160$K)\cite{Mitterdorfer2014,Mate2012,Stevenson1999}\footnote{The anneal temperature (200K) was measured at the base of the sample holder. In theory the limited heat conductivity could lead to temperature gradients and a different temperature at the surface of the sample. However, the fact that the sample (with a volume of $\approx 10$ mm$^3$ and a surface area of $\approx 10$ mm$^2$ evaporated completely within about half an hour agrees with the known vapour pressure at 200K, and rules out a scenario where regions of the sample have remained colder than the crystallization temperature.} . Consequently, if indeed the relative intensity of the faster relaxing component was related to the number of molecules located at, or close to, the surface of pores, one could expect the relative intensity of this component to vanish or at least significantly decrease after annealing, similarly to what was observed in the dielectric measurements\cite{Johari1991}.

On the other hand, the sample we studied was relatively thick ($\approx$1mm) and consequently also the growth rate was substantially larger than the rates used in the measurements which followed the dynamics of pore collapse. Furthermore, it has been shown that faster deposition enhances the porosity and leads to a slower sintering process\cite{Mitterdorfer2014}, which raises the question whether the pores in our sample could remain intact even when annealing to temperatures where significant sample desorption already takes place?

In order to further investigate this hypothesis, we measured ESEM images from vapor deposited ices samples. Cartwright et al. used ESEM to study ASW films which were grown at very fast deposition rates\cite{Cartwright2010}. The images revealed remarkably complex (and beautiful) meso-scale morphologies which strongly depended on the deposition rates and surface temperatures. These observations motivated us to perform complementary ESEM measurements of the evolution of the meso-scale morphology when annealing cold-grown vapor deposited ice samples. 

Figures 4 and 5 show ESEM images of ice samples grown using a fixed deposition rate provided by a water partial pressure of P = $5\cdot 10^{-3}$ Torr (equivalent to a molecular flux of $3.5 \cdot 10^{17}$ cm$^{-2}$ sec$^{-1}$). At the end of the deposition, the microscope was pumped down to its base pressure (P$<10^{-5}$ torr) where images were acquired.

In figure 4 we show ESEM images of two films, one of which was grown at 80K (left panel) and the other at 180K. The sample which was grown at a lower temperature has a ``cauliflower'' type morphology, characterized by a high surface-area strongly textured interface. In contrast, a sample grown at a much higher temperature (right panel), is characterized by micron-sized faceted prismatic domains, suggesting some degree of crystallinity, and a significantly lower surface area. 

Figure 5 shows the morphology of a sample which was grown at a low temperature (80K) and annealed at 180K. The left and the right panels of figure 5 show the same region of the sample before and after annealing it for 10 minutes at 180K. The complex internal microstructure, which characterizes the cold grown samples, does not show any signs of sintering on the micron scale, as could be perhaps expected at this relatively high temperature. Instead what can be seen from comparing the two images, is a widening of the cracks and a reduction of the size of the micron sized domains due to significant desorption. Thus, desorption of molecules to the gas phase seems to be more efficient than the transport required for sintering the sample and removing the high surface area microporous morphology seen in the image. 

It is important to note, that ESEM images characterize the mesoscopic (micron) scale, whereas NMR is sensitive to the immediate molecular environment, hence, care should be taken when comparing the two. Nevertheless, if the morphologies of these different length scales are related, then the ESEM results support the possibility that the fast relaxation component in the NMR measurements is linked to molecules at the pore surfaces, as the notion that efficient sintering and pore collapse should take place before evaporation, does not seem to apply to the samples we studied. While at first sight, this finding might seem to contradict previous observations (e.g. \cite{Mitterdorfer2014,Mate2012, Johari1991}) we believe that the trend observed in the past\cite{Mitterdorfer2014}, where pore collapse can be delayed to hotter temperatures when growing the films faster, is further extended due to our even faster deposition rates, leading to a situation where the sample evaporates well before effective sintering can take place.

\begin{figure}
\includegraphics[width=170mm]{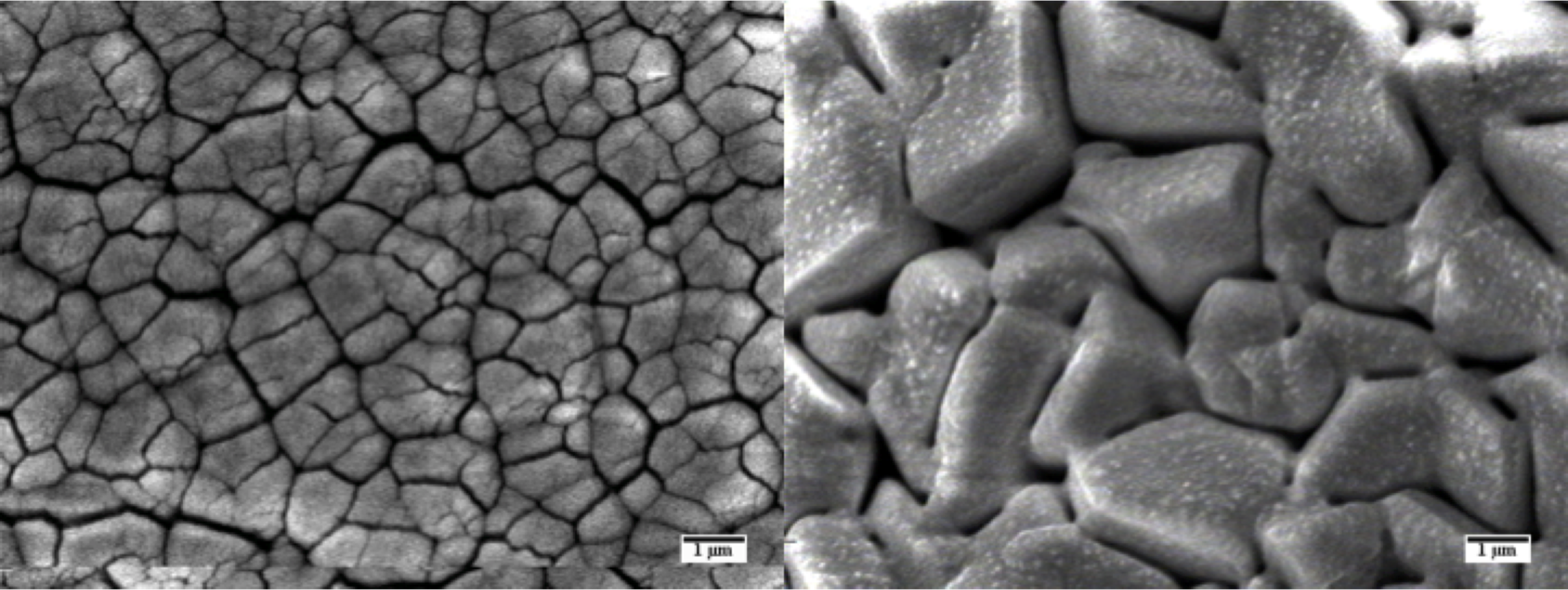}
\caption{\label{cap:ESEM}ESEM images of vapor deposited ice samples grown at 80 K (left image) and at 180K (right image). The morphology on the micron scale changes dramatically, from high-surface-area ``cauliflower type'' structure to relatively smooth faceted prismatic domains.}
\end{figure}

\begin{figure}
\includegraphics[width=170mm]{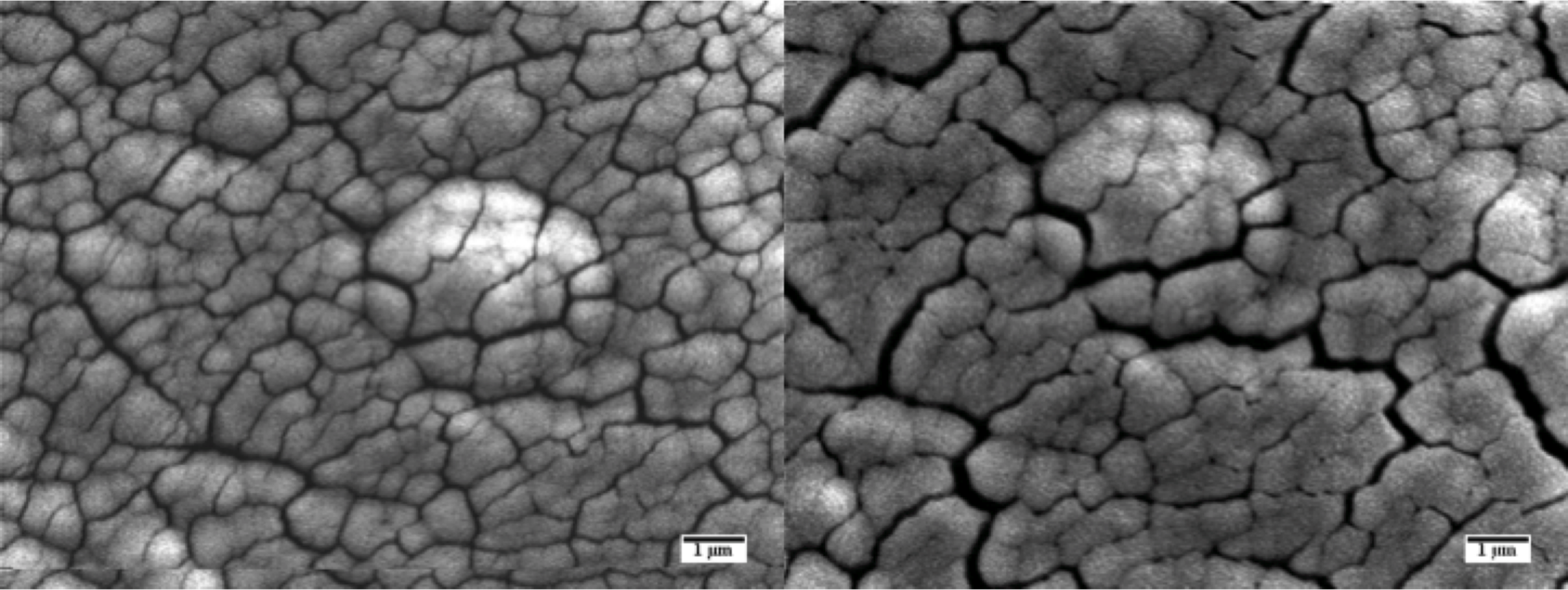}
\caption{\label{cap:ESEM2}ESEM images of a vapor deposited ice sample grown at 80 K before (left image) and after (right image) annealing at 180K for 10 minutes. The only significant change which can be seen to have happened is desorption, reflected in the widening of the cracks and a reduction in the size of the ice micro-structures.}
\end{figure}

\section{Summary and conclusions}

The apparatus presented in this work allows in-situ NMR measurements of vapor deposited ice samples. The spin-lattice relaxation curves we measured are complex and depend strongly on the growth conditions of the sample. Samples grown below the amorphous to crystalline transition temperature relax to their spin equilibrium on two distinct time scales, differing by more than 3 orders of magnitudes. In contrast, when growing the sample above the crystallization temperature, we observe only one slow relaxation process. 

The slow relaxation component seen in the cold-grown samples becomes faster when heating the samples within the 52K to 140K range, suggesting that this relaxation process arises from temperature activated motion, such as vibrations, librations, the diffusion of defects within the network, or some other form of dynamics. Annealing the sample at 200K slows down this relaxation component, indicating that either an annealing effect or the transition from amorphous to crystalline (or both) makes the relaxation mechanism less effective. An elongation of the spin lattice relaxation times due to crystallization, resembles previously observed trends where $^1$H spin-lattice relaxation times were shown to be shorter in pressure induced amorphous ice (LDA) in comparison with crystalline ice\cite{Ripmeester1992}, observations which are consistent with the reduction of diffusivity when transforming from an amorphous form to crystalline ice\cite{Smith1999}.

In contrast, the fast relaxing component which is only seen in the cold-grown samples, does not seem to be affected significantly by heating the sample and its relative contribution does not change even after annealing at 200K. Thus, the sample evaporates before this particular characteristic of cold-grown samples can be erased by an annealing or sintering process. One mechanism which could produce a significantly enhanced relaxation rate for cold grown samples is the existence of a significant population of water molecules located close to the pores. 

The fact that this fast relaxation is observed to persist even after annealing the sample to high temperatures, combined with the fact that previous studies have shown pore collapse well below the amorphous to crystalline transition temperature \cite{Mitterdorfer2014,Mate2012, Johari1991} seems at first to raise doubts whether indeed the pores are related to the observed fast relaxation. However, the ESEM measurements we present show that the micro-porous structures of cold-deposited ices can remain intact at elevated temperatures and that evaporation is more efficient then sintering. We believe the reason that the samples we studied (with NMR and ESEM) behave differently than those studied in the past using different experimental methods, is related to the faster deposition rates used in this study. Further measurements of samples grown using a smaller deposition flux, requiring some modification of the experimental setup, are needed to check this hypothesis and further characterize the mechanism underlying the spin-lattice relaxation process in vapor deposited ice samples. 

\section{Acknowledgements}

This work was funded by the German-Israeli Foundation for Scientific Research, BSF Grant 2010095, ISF Grant 755/16 and the European Research Council under the European Unions seventh framework program (FP/2007-2013)/ERC Grant 307267. P.A. acknowledges support from NSERC, FRQNT, and CFI. The authors would like to thank Prof. Shimon Vega, Dr. Akiva Feintuch, Prof. Franz. Fujara, and Prof. John Ripmeester for their valuable advice and support.

\bibliographystyle{unsrt}

%\bibliographystyle{unsrt}

%\bibliography{langevin_refs}

\end{document}